\documentclass[conference]{IEEEtran}
\IEEEoverridecommandlockouts
\usepackage{amsmath,amssymb,amsfonts}
\usepackage{algorithmic}
\usepackage{graphicx}
\usepackage{textcomp}
\usepackage{xcolor}

\usepackage[indentfirst=false,font={itshape},begintext=``,endtext='']{quoting}
\usepackage[square,numbers]{natbib}
\usepackage{subcaption}
\usepackage{hyperref}
\usepackage{csquotes}  
\usepackage{tablefootnote}
\usepackage{booktabs}
\usepackage{balance} 
\usepackage{amssymb}

\makeatletter

\newcommand{\ygg@basicalert}[2]{\textcolor{red}{\fbox{\bfseries\sffamily\scriptsize#1}{\sf\small$\blacktriangleright$\textit{#2}$\blacktriangleleft$}}}
\newcommand{\YANN}[1]{\ygg@basicalert{YANN}{#1}}
\newcommand{\GABRIEL}[1]{\ygg@basicalert{GABRIEL}{#1}}

\def\BibTeX{{\rm B\kern-.05em{\sc i\kern-.025em b}\kern-.08em
    T\kern-.1667em\lower.7ex\hbox{E}\kern-.125emX}}
    
\begin{document}

\title{An Exploratory Approach for Game Engine Architecture Recovery}

\author{\IEEEauthorblockN{1\textsuperscript{st} Gabriel C. Ullmann}
\IEEEauthorblockA{\textit{Concordia University} \\
Montreal, Quebec, Canada \\
g\_cavalh@live.concordia.ca}
\and
\IEEEauthorblockN{2\textsuperscript{nd} Yann-Ga\"el Gu\'{e}h\'{e}neuc}
\IEEEauthorblockA{\textit{Concordia University} \\
Montreal, Quebec, Canada \\
yann-gael.gueheneuc@concordia.ca}
\and
\IEEEauthorblockN{3\textsuperscript{rd} Fabio Petrillo}
\IEEEauthorblockA{\textit{École de Technologie Supérieure} \\
Montreal, Quebec, Canada \\
fabio.petrillo@etsmtl.ca}
\and
\hspace{3cm}\IEEEauthorblockN{4\textsuperscript{th} Nicolas Anquetil}
\IEEEauthorblockA{
    \hspace{3cm}\textit{Univ. Lille, CNRS, Inria, Centrale Lille,} \\ \hspace{3cm}\textit{UMR 9189 - CRIStAL} \\
    \hspace{3cm} {Lille, France} \\
    \hspace{3cm} {nicolas.anquetil@inria.fr}
}
\and
\IEEEauthorblockN{5\textsuperscript{th} Cristiano Politowski}
\IEEEauthorblockA{\textit{École de Technologie Supérieure} \\
Montreal, Quebec, Canada \\
cristiano.politowski@etsmtl.ca}

}

\maketitle

\begin{abstract}
Game engines provide video game developers with a wide range of fundamental subsystems for creating games, such as 2D/3D graphics rendering, input device management, and audio playback. Developers often integrate these subsystems with other applications or extend them via plugins. To integrate or extend correctly, developers need a broad system architectural understanding. However, architectural information is not always readily available and is often overlooked in this kind of system. In this work, we propose an approach for game engine architecture recovery and explore the architecture of three popular open-source game engines (Cocos2d-x, Godot, and Urho3D). We perform manual subsystem detection and use Moose, a platform for software analysis, to generate architectural models. With these models, we answer the following questions: Which subsystems are present in game engines? Which subsystems are more often coupled with one another? Why are these subsystems coupled with each other? Results show that the platform independence, resource management, world editor, and core subsystems are frequently included by others and therefore act as foundations for the game engines. Furthermore, we show that, by applying our approach, game engine developers can understand whether subsystems are related and divide responsibilities. They can also assess whether relationships among subsystems are appropriate for the game engine.
\end{abstract}


\section{Introduction}
\label{sec:introduction}

Game engines are tools to facilitate game development. They are systems composed of subsystems that interact with several audio, video, network, storage, and user interface devices. Besides making these subsystems work together with high performance, developers often integrate them with other software or extend them using plugins. However, to make the right choices, developers first need to understand the game engine architecture they are working with and how its subsystems relate to one another: ``[a] prerequisite for integration and extension is the comprehension of the software. To understand the architecture, we should identify the architectural patterns involved and how they are coupled.'' \cite{agrahari_2021}.

Even though these are relevant concerns on both game engine and game development, the topics of game engine subsystems coupling and architectural models have not been explored extensively. The literature mostly focuses on the implementation of specific game engine subsystems and not on their design and integration in architecture. In the context of graphics, for example, ``there are a lot of sources of very good information from research to practical jewels of knowledge. However, these sources are often not directly applicable to production game environments or suffer from not having actual production-quality implementations.'' \cite[p.~xiv]{gregory_game_2018}.

Moreover, popular game engines, such as Unity \footnote{\url{https://unity.com/solutions/game}}, are closed source, which makes it harder for developers to study them \cite{ullmann_anatomy_2022}. When developers have no access to architectural models, the software understanding process is mostly based on trial and error: ``the only way for a developer to understand the way certain components work and communicate is to create his/her own computer game engine.'' \cite{srsen_2021}. 

In this work, we propose an approach for creating architecture models of game engines. Given a game engine, we cluster its source-code files into subsystems and create an \textit{include} graph that holds the representation of the dependency relationships between these files. After creation, we check this graph for inconsistencies, resolving manually any \textit{include} paths that could not be resolved automatically by our graph generator. Finally, we use this graph to generate an intermediate model that can be loaded into Moose 10, a platform for software analysis \footnote{\url{https://moosetechnology.org}}. Using this platform, we generate an architectural model that allows us to visualise the relationships between subsystems. With such a model we can answer the following research questions:

\begin{itemize}
\item \textbf{RQ1}: Which subsystems are present in game engines? 

\item \textbf{RQ2}: Which subsystems are more often coupled with one another?

\item \textbf{RQ3}: Why are these subsystems coupled with each other?
\end{itemize}

We apply our approach to three popular open-source game engines: Cocos2d-x \footnote{\url{https://github.com/cocos2d/cocos2d-x}}, Godot \footnote{\url{https://github.com/godotengine/godot}}, and Urho3D \footnote{\url{https://github.com/urho3d/urho3d}}. Our results show that game engines share the same set of base subsystems, even though some responsibilities are grouped differently on each system. For example, files from \textit{Profiling and Debugging} and \textit{Human Interface Devices} subsystems can be either placed in separate folders or merged together in a single ``core'' folder. We also observe that the \textit{Platform Independence Layer}, \textit{Resource Management}, \textit{World Editor}, and \textit{Core} subsystems are frequently included by others and therefore are foundational for game engines. Moreover, we observe that the \textit{Core} subsystem frequently includes graphic-related subsystems to initialise these subsystems and access debugging information. We show and discuss more examples of these and other occurrences. 

We show that, by applying our approach, game engine developers can understand whether subsystems are related and how responsibilities are/should be divided. Also, they can assess if these relationships help the subsystems to fulfil their responsibilities in the overall system. We intend to apply this approach to a larger set of game engines and subsystems. We also want to explore other relationships and metrics, such as cohesion and complexity, and their effect on game engine understanding and maintainability.

The remainder of the paper is organised as follows: Section \ref{sec:related-work} presents related work on game engines, architectural recovery, and coupling. Section \ref{sec:method} provides a description of our game engine analysis approach. Section \ref{sec:results} shows and discusses the architectural models resulting from applying our approach to three game engines. Section \ref{sec:threats} presents threats to validity and Section \ref{sec:conclusion} concludes with future work.

\section{Related Work}
\label{sec:related-work}
Works compared game engine aspects such as ease of use \cite{dickson_experience-based_2017}, available subsystems and target platforms \cite{mishra_comparison_2016}, and suitability for a given platform \cite{pattrasitidecha_comparison_2014} or game genre \cite{pavkov_comparison_2017}. These comparisons are all tabular, listing game engines in one axis and relating them to subsystems in another. They are organized in this way to determine which game engines have the largest subsystem count, which subsystems are more commonly implemented and what benefit they bring to developers. But while discussing function, these works do not mention the architectural structures which support functionality. In this paper, we present an approach to obtain architectural models from game engines, which may be useful for both game engine developers and game developers who want to integrate or extend game engines.

Outside of game engines, the recovery and study of software architecture models have been applied to different software, such as the Linux kernel \cite{bowman_linux_1999} and the IBM SQL/DS database \cite{wong_structural_1995}. In their work, \cite{bowman_linux_1999} present a model showing the relationships among Linux subsystems. This visualisation, which they called ``conceptual architecture'', helped them ``understand the volume of detail in the implementation'' and view ``relationships between subsystems that are \enquote{meaningful} to developers''. Similarly, in this work, we propose an approach to obtain game engine architectural models by detecting and relating their subsystems. We present our analysis of subsystem coupling with an architectural model generated by Moose, which represents each subsystem as a node in a graph, with edges between them representing dependency relationships. 

Many software quality metrics exist and we choose coupling because coupling among subsystems directly impacts understanding and maintainability \cite{wilkie_coupling_2000}, which are important to consider because video games are ``complex, emergent systems that are difficult to design and test'' \cite{lewis_what_2010}. Also, ``the more scattered the [game engine] feature implementations are in the architecture, the more likely it is that they get tangled with features to be added.'' \cite{guana_building_2015}.

In previous work, we compared the dynamic call graphs of Godot and Urho3D to understand the similarities and differences between their initialisation processes and division of responsibilities at the class level \cite{ullmann_anatomy_2022}. In this work, we analyse the architecture at the  subsystem level, observing the static dependencies among files and not their method calls. We analyse the same two engines, Godot and Urho3D, and add a third one, Cocos2d-x.

\section{Approach}
\label{sec:method}

\begin{figure}[ht]
\center
\includegraphics[width=\columnwidth]{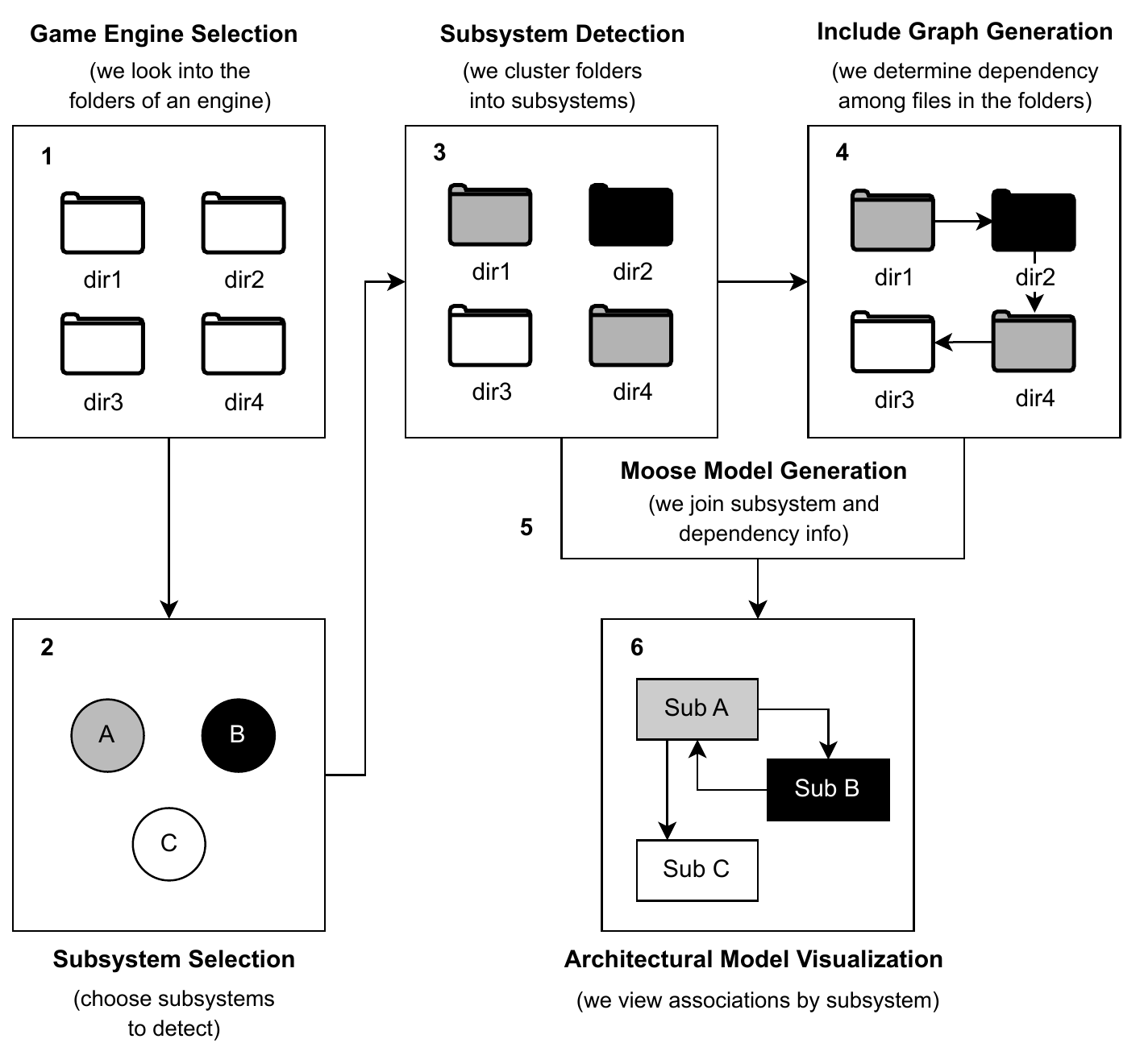}
\caption{Steps of our game engine analysis approach.}
\label{fig:diagram-approach}
\end{figure}

We divide our approach into six steps (\autoref{fig:diagram-approach}): system selection, subsystem selection, subsystem detection, generation of the \textit{include} graph, Moose model generation, and architectural model visualisation. We partially automate our approach. Data and scripts are available on GitHub \footnote{\url{https://github.com/gamedev-studies/game-engine-analyser}}. In this section, we explain our steps and choices in detail.

\begin{enumerate}
\item \textbf{System selection}: We chose the game engines to study using GitHub. This step was done manually.

\item \textbf{Subsystem selection}: We chose game engine subsystems from the game engine development literature. This step was done manually.

\item \textbf{Subsystem detection}: For each selected game engine, we clustered files and folders into the selected subsystems. This process was done manually by a single person in this paper but could be done by multiple people and the results could be combined by consensus.

\item \textbf{Include graph generation}: We generated an \textit{include} graph that encompasses all the files of a selected game engine. This step was done automatically with a dedicated tool, described in Section \ref{sec:include-graph-generation}.

\item \textbf{Moose model generation}: We used the data obtained from Steps 3 and 4 to generate a model that can be loaded into the Moose software analysis platform. This step was done semi-automatically with dedicated scripts.

\item \textbf{Architectural model visualisation}: We loaded each game engine model into Moose and then used its ``Architectural map'' visualisation to generate a visual representation of the \textit{include} graph and subsystems. This representation is an architectural model. This step was done semi-automatically with dedicated scripts.
\end{enumerate}

\subsection{System Selection}
\label{sec:sys-selection}

We searched for and selected game engine repositories in GitHub. We defined the following criteria for selecting game engine repositories:  

\begin{itemize}
\item \textbf{Open source}: the repository must be publicly available and unarchived.

\item \textbf{Game engine}: the repository must be tagged with the ``game engine'' tag.

\item \textbf{Written in C++}: the repository must contain mostly C++ code because it is the most used programming language for game engine development \cite{politowski_engines_2021}.

\item \textbf{General purpose}: the game engine must enable the creation of games of diverse game genres. We made this choice so our analysis reflects the architectural choices and needs of a broad range of games and is useful for a large set of game developers.
\end{itemize}

In our search result, we selected the top three engines with the highest sum of forks and stars and with the lowest numbers of header files and folders because our approach is partly manual. We thus obtained three game engines: Cocos2d-x, Godot, and Urho3D, as shown in Table \ref{tab:engine-selection}.

\begin{table*}[ht]
\centering
\caption{Open-source game engines in GitHub which match our criteria.}
\label{tab:engine-selection}
\begin{tabular}{@{}lllrrr@{}}
\toprule
\textbf{Game Engine} & \textbf{Branch}                                              & \textbf{Commit} & \multicolumn{1}{l}{\textbf{Forks + Stars}} & \multicolumn{1}{l}{\textbf{Headers}} & \multicolumn{1}{l}{\textbf{Folders}} \\ \midrule
Cocos2d-x          & HEAD-$>$v4                         & 90f6542cf7      & 23,300                                      & 1,089                                      & 642                                  \\
Godot              & HEAD-$>$3.4;tag:3.4.5-stable      & f9ac000d5d      & 59,200                                      & 2,748                                      & 1,022                                 \\
Urho3D             & HEAD-$>$master                     & feb0d90190      & 4,956                                       & 3,446                                      & 1,546                                 \\
\bottomrule
\end{tabular}
\end{table*}

\subsection{Subsystem Selection}
\label{sec:sub-selection}
We used the 15 subsystems described in the ``Runtime Engine Architecture'' proposed by \cite[p.~33]{gregory_game_2018}. We chose these subsystems because they are well-known in the game engine development community. While \citeauthor{gregory_game_2018} uses the terms ``component'', ``module'' and ``subsystem'' interchangeably, we chose to use the word ``subsystem'' to describe a group of files and folders that contain code for a given system functionality.

\citeauthor{gregory_game_2018} states that ``all commercial game engines have some kind of world editor tool'', which is ``a tool that permits game world chunks to be defined and populated'' \cite[p.~857]{gregory_game_2018} so we added a subsystem, which we called ``World Editor'', because it directly impacts game developers' work.

While the description of the responsibilities encompassed by each subsystem is beyond the scope of this paper, we summarise their definitions in the following. We also assign them a 3-letter code, which we use in the rest of this paper. 

\begin{enumerate}
\item \textbf{Audio (AUD)}: manages audio playback and effects.
\item \textbf{Core Systems (COR)}: manages engine initialisation, contains libraries for math, memory allocation, etc.
\item \textbf{Profiling and Debugging (DEB)}: manages performance stats, debugging via in-game menus or console.
\item \textbf{Front End (FES)}: manages GUI, menus, heads-up display (HUD), and full-motion video playback.
\item \textbf{Gameplay Foundations (GMP)}: manages the game object model, scripting and event/messaging system.
\item \textbf{Human Interface Devices (HID)}: manages game-specific input interfaces, physical I/O devices.
\item \textbf{Low-Level Renderer (LLR)}: manages cameras, textures, shaders, fonts, and general drawing tasks.
\item \textbf{Online Multiplayer (OMP)}: manages match-making and game state replication.
\item \textbf{Collision and Physics (PHY)}: manages forces and constraints, rigid bodies, ray/shape casting.
\item \textbf{Platform Independence Layer (PLA)}: manages platform-specific graphics, file systems, threading, etc.
\item \textbf{Resources (RES)}: manages the loading and caching of game assets, such as 3D models, textures, fonts, etc.
\item \textbf{Third-party SDKs (SDK)}: enables interfacing with DirectX, OpenGL, Vulkan, Havok, PhysX, STL, etc.
\item \textbf{Scene graph/culling optimizations (SGC)}: computes spatial hash, occlusion, and level of detail (LOD).
\item \textbf{Skeletal Animation (SKA)}: manages animation state tree, inverse kinematics (IK), and mesh rendering.
\item \textbf{Visual Effects (VFX)}: enables light mapping, dynamic shadows, particles, decals, etc.
\item[$+$] \textbf{World Editor (EDI)}: visual game world-building.
\end{enumerate}

\subsection{Subsystem Detection}
\label{sec:sub-detection}

In this step, we clustered all folders in each repository into the selected subsystems. When deciding which subsystem best suits a given folder, we considered all available information about the folder: name, contents, documentation, and source code. We show an example of this decision process in \autoref{tab:sub-id-cocos2dx}.

\begin{table}[ht]
\centering
\caption{Subsystem detection example for Cocos2d-x.}
\label{tab:sub-id-cocos2dx}
\begin{tabular}{@{}p{3.7cm}p{4.2cm}@{}}
\toprule
\textbf{/cocos/editor{\textendash}support/spine}  & \\ \midrule 
\textbf{Subsystem detected:}          & \\
\textbf{1) By folder name?}           & No, there is no subsystem called \textit{spine}. From the names of the files it contains, we can infer it is related to animation (SKA), but we need more data to confirm. \\
\textbf{2) By parent folder name?}    & No, the folder \textit{editor-support} might be related to EDI, but we need more data to confirm. \\
\textbf{3) By documentation? }         & Yes, according to docs: ``Skeletal animation assets in Creator are exported from Spine''. \tablefootnote{\url{https://docs.cocos.com/creator/manual/en/asset/spine.html}}\\
\textbf{4) By code?}                  & No need to look at the code, subsystem detected on step 3. \\ \midrule 
\textbf{Conclusion}                   & SKA \\ \bottomrule
\end{tabular}
\vspace{-0.40cm}
\end{table}

\subsection{Include Graph Generation}
\label{sec:include-graph-generation}

We generated an \textit{include} graph for each repository using a script created by Irving\footnote{\url{https://www.flourish.org/cinclude2dot}}. This script generates a graph file in the DOT language, which contains each file's absolute path and its \textit{include} relations with other files.

\begin{figure}[ht]
\center
\includegraphics[width=\columnwidth]{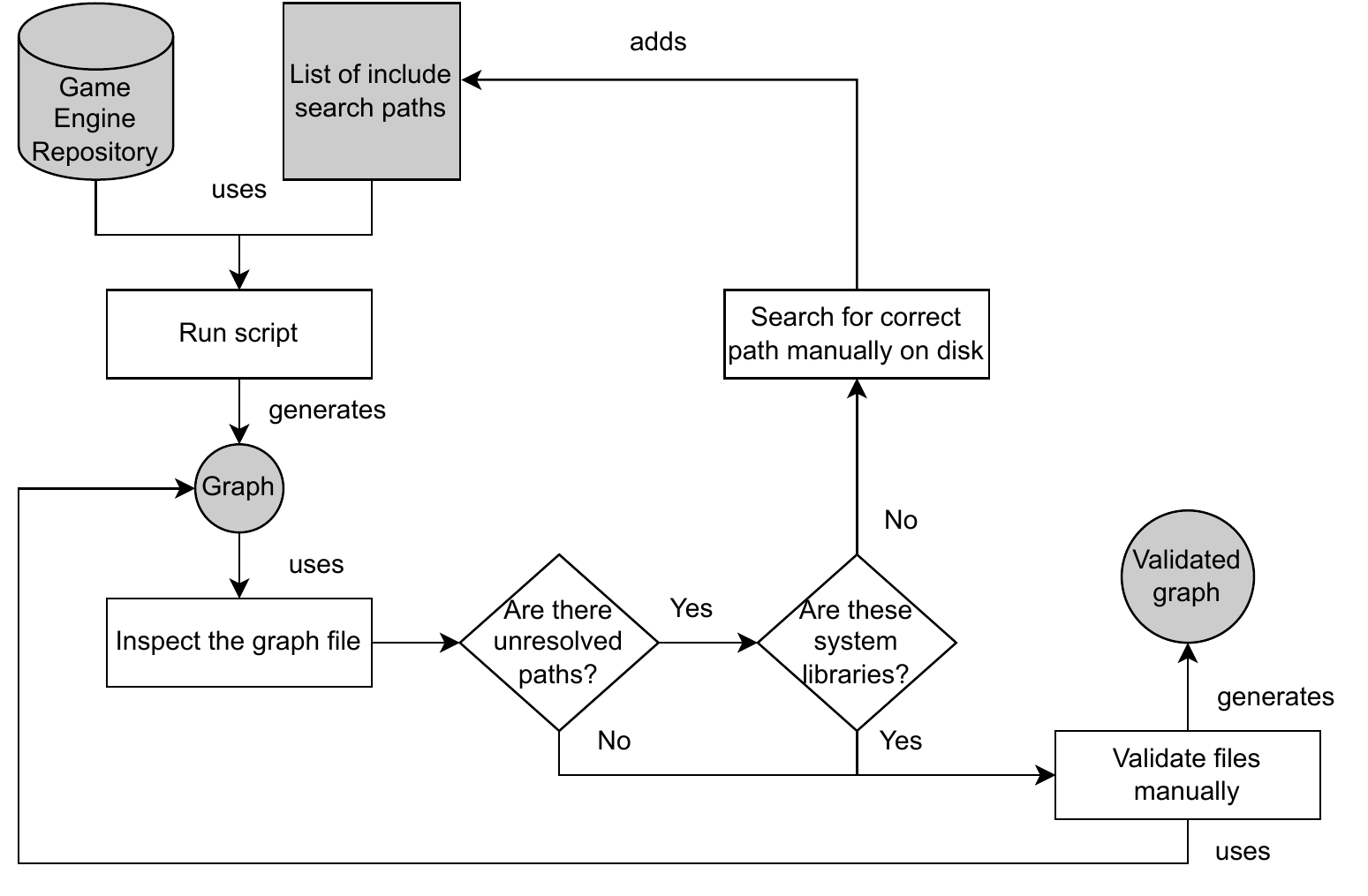}
\caption{Steps of \textit{include} graph generation.}
\label{fig:diagram-includes}
\end{figure}

The script attempts to resolve each relative \textit{include} path into an absolute path. If the resolution fails, the script stores the path in a file. We then iterate over this file and resolve the paths manually. After resolution, we added the path to the script list of search paths and ran the script again (\autoref{fig:diagram-includes}).

Some \textit{include} paths remained unresolved because they referred to system or OS-specific libraries (e.g., \textit{stdio.h}, \textit{windows.h}). These headers are external to the game engine repositories so their absence does not influence our results. 

\subsection{Moose Model Generation}
\label{sec:moose-model-generation}

We used Moose 10, a platform for software analysis, to analyse the \textit{include} graphs. We chose Moose because it generates visualisations and is built on top of Pharo\footnote{\url{https://pharo.org/}}, a Smalltalk development environment that gives us the flexibility to customize existing tools and also write our own.

We wrote a Smalltalk program that takes as input a list of game engine files/folders, their assigned subsystem and \textit{include} graph files. The union of these two data sources happens in Step 4 of \autoref{fig:diagram-approach}. The result of this union is a Moose model, which is a Pharo object containing source-code entities (files and folders) and \textit{include} relations. We then used Moose and Pharo methods to analyse these models, such as counting the number of entities in a model, filtering them by name and type, or displaying them with the ``Architectural map'' visualisation, as described in \ref{sec:arch-map-generation}.

\subsection{Architectural Model Visualisation}
\label{sec:arch-map-generation}

Moose provides an ``Architectural map'' to visualise entities and relationships in Moose models. It is a directed graph where each model entity is a node, and each \textit{include} relationship is an edge. This graph is an architectural model.

The ``Architectural map'' visualisation is interactive and allows us to group files/folders by tags, identified by names and colours. We created one tag for each selected subsystem and assigned one tag to each entity in each model. The tag creation and assignment process was semi-automated. We wrote a script to gather all the folder names and their respective subsystem names (see Section \ref{sec:sub-detection}) from a CSV file and then create the tags. We created the CSV files manually.

Once we created all tags in Moose for each game engine, we selected all entities at the root folder, propagated them to the data bus, opened the ``Architectural map'' and finally selected all tags and relationships (called ``associations'' in Moose) for visualisation. We describe and comment on the results next.

\section{Results}
\label{sec:results}

\begin{figure*}[!t]
    \centering
    \begin{subfigure}[b]{0.49\textwidth}
        \includegraphics[width=\textwidth]{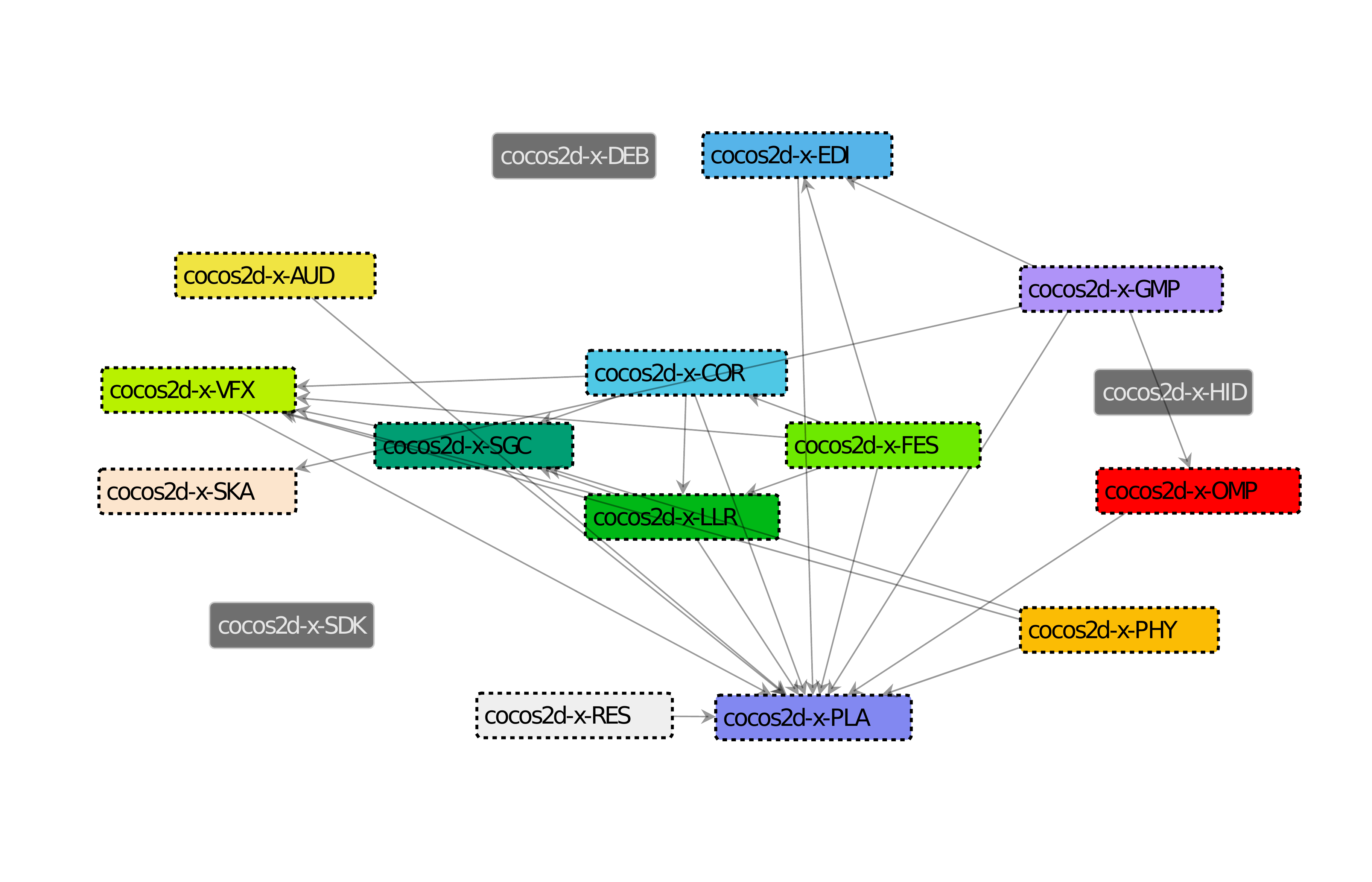}
        \caption{Cocos2d-x}
      \end{subfigure}
      \begin{subfigure}[b]{0.49\textwidth}
        \includegraphics[width=\textwidth]{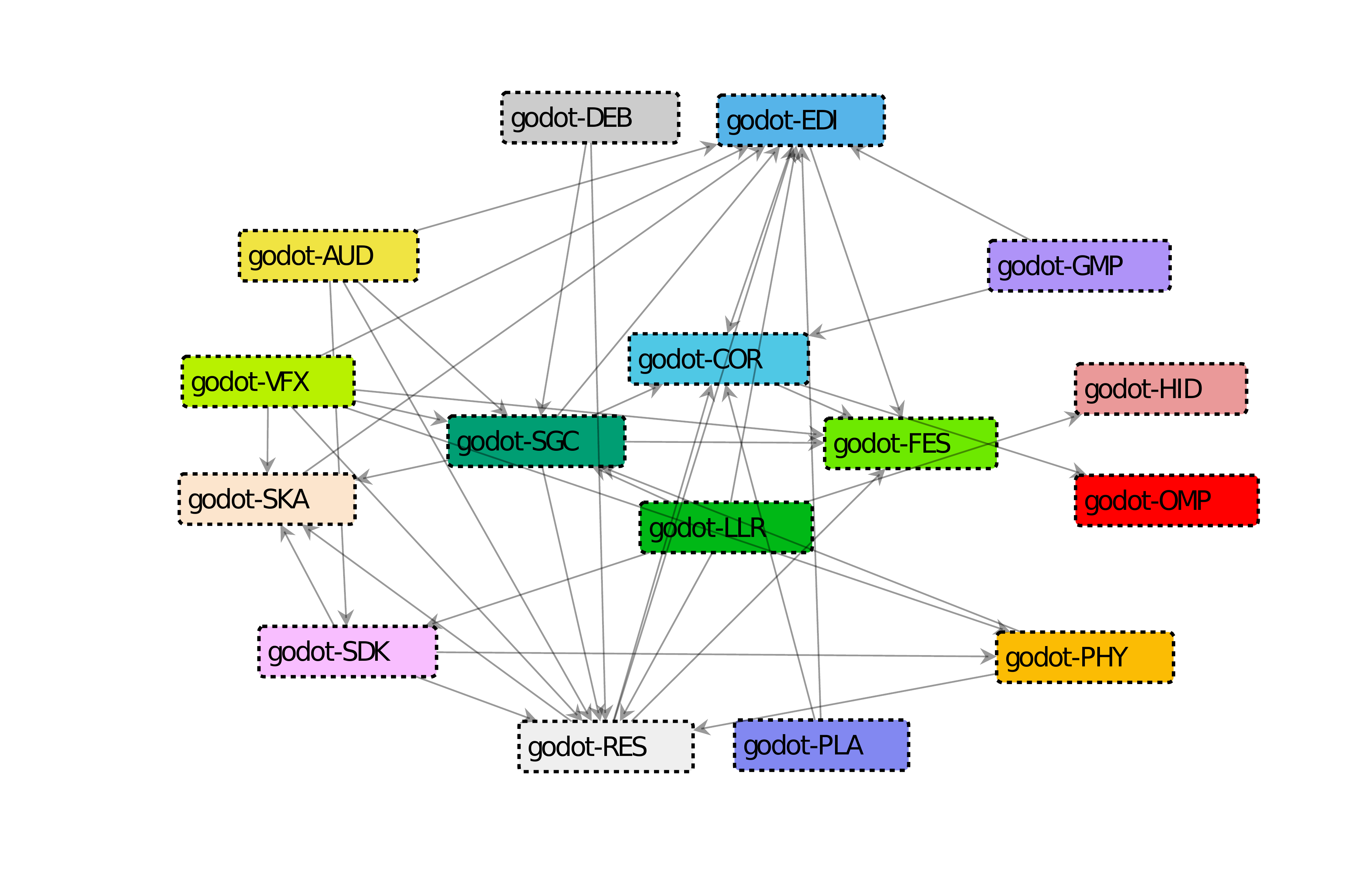}
        \caption{Godot}
      \end{subfigure}
      \begin{subfigure}[b]{0.49\textwidth}
        \includegraphics[width=\textwidth]{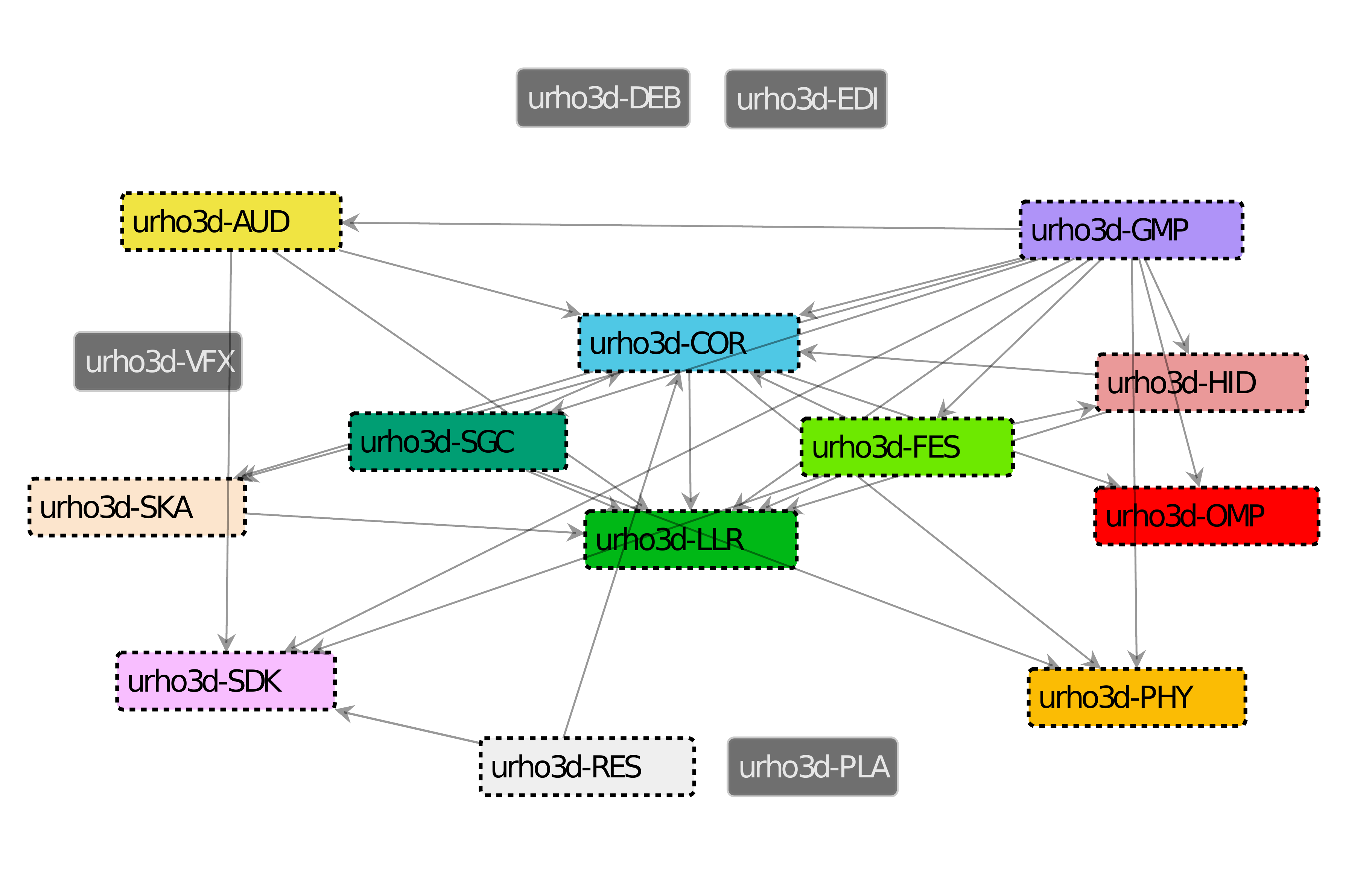}
        \caption{Urho3D}
      \end{subfigure}
      \caption{Game engine architectural models generated with Moose 10.}
      \label{fig:arch-maps}
\end{figure*}

The application of our approach on the three selected game engines produces three architectural models showing the similarities and differences among subsystems and their relations, shown in \autoref{fig:arch-maps}. We discuss the architectural models and answer our RQs.

\subsection{RQ1: Detected Subsystems}
\label{sec:discussion-subs}

In the architectural models shown in \autoref{fig:arch-maps}, all subsystems are identified by coloured boxes named with the abbreviations in Section \ref{sec:sub-selection}. Subsystems that were not detected are represented by a dark grey box with light grey text. The four subsystems placed in the centre of the map are the ones most often included by others. We placed the subsystems manually in a circular pattern, organized alphabetically from left to right.

We identified 12 subsystems in Urho3D, 13 in Cocos2d-x and all 16 in Godot. While this may imply that Godot is more feature-complete than its counterparts, an undetected subsystem does not mean that its features are totally absent from the game engine. For example, we found out that DEB subsystem functionality is not always centralized in a file/folder, but rather spread among different subsystems. Concretely, we observed in Urho3D that both files \textit{Renderer.h} and \textit{DebugRenderer.h} inside the \textit{Graphics} folder, clustered into the LLR subsystem. LLR has its own debugging utilities, no system-wide debug utilities exist.

We observed that some subsystems are present but their files are in a \textit{base} or \textit{core} folder, mixed with other, unrelated files. Our clustering was mainly based on folder names so these files were clustered together, even though they are not functionally related. For example, we confirmed that Cocos2d-x has HID functionality by inspecting its source files, which is backed by its documentation\footnote{\url{https://docs.cocos2d-x.org/cocos2d-x/v4/en/event\_dispatcher/keyboard.html}}. However, the files that contain HID-related classes are in a folder called \textit{base}, which was clustered in the COR subsystem.

Urho3D presents several other cases where one of the selected subsystems is encompassed by another. Code for MacOS compatibility, a PLA subsystem responsibility, can be found in folders such as \textit{IO} (COR subsystem) and \textit{ThirdParty} (SDK subsystem). VFX subsystem features are in files under \textit{Graphics}, along with \textit{Renderer.h}. 

Finally, we observed cases where a subsystem was not detected because its files were in a separate folder or repository. In Urho3D, for example, the editor is represented by a single AngelScript file, \textit{Editor.as}, which is in \textit{/bin/Da\-ta/Scripts} along with other code examples and away from \textit{/Source} in which most subsystem code resides. This subsystem is not written in C++ while, in this work, we only considered C++ files. In Cocos2d-x, third-party libraries are placed in a separate repository\footnote{\url{https://github.com/cocos2d/cocos2d-x-3rd-party-libs-bin}}, which we excluded from our analysis. 

\subsection{RQ2 and RQ3: Coupling Among Subsystems}
\label{sec:discussion-coupling-subs}

The architectural models in \autoref{fig:arch-maps} go beyond showing the existence of a set of subsystems or comparing their implementation with a theoretical model. Indeed, they are directed graphs whose incoming and outgoing edges we can study to understand subsystem relationships. We can also answer questions to explore and validate the architectures, such as ``Should subsystem A include files from B? If so, why?''.

Of the three architectural models, Godot is noticeably the densest one not only due to the number of subsystems but also edges, a total of 40. Urho3D is the second most coupled game engine, with 24 edges, followed by Cocos2d-x with 21. These numbers hint at the sizes and complexities of the selected game engines. Yet, they do not provide specific insights. In the following, we analyse the architectural models in-degree and out-degree and compare the frequencies of coupling between subsystems across game engines. 

\subsubsection{In-degree and Out-degree}
\label{sec:discussion-in-out-degree}

In graph theory, the number of incoming edges of a node is called in-degree, and the number of outgoing edges is called out-degree. We noticed high in-degree nodes in all game engines, and so we decided to compute a node count for each node on each game engine to understand which subsystems support others.

On the bottom of the Cocos2d-x architectural model, we observe a node with a very high in-degree, the PLA subsystem, with 11 incoming and zero outgoing edges: 11 subsystems depend on PLA because the folder \textit{cocos/platform} contains many files with utility classes used throughout the system. Some examples are \textit{CCFileUtils.h} (file management), \textit{CCImage.h} (image loading), \textit{CCPlatformConfig.h} (checks for cross-platform compatibility), and \textit{CCLuaEngine.h} (scripting engine). These files could be clustered into their specific subsystems but many account for OS-specific considerations (e.g., different line-breaking characters, asset root folders, etc.) so they belong to the PLA subsystem.

The RES subsystem has the highest in-degree in Godot, with seven incoming and four outgoing edges. It is followed closely by EDI, with eight incoming and two outgoing edges. These subsystems have high in-degrees, similar to what we observed in Cocos2d-x. The RES subsystem contains files with classes representing game entities from many subsystems, such as \textit{audio\-\_stream\-\_sample.h} (AUD), \textit{dynamic\_font.h} (FES), \textit{phy\-sics\_\-material.h} (PHY/LLR), and \textit{animation.h} (SKA). 

In Godot, most files that depend on its EDI subsystem are custom engine modules, located in \textit{godot/modules}. These modules extend editor functionality for different subsystems, e.g., GMP (\textit{mo\-dules/gd\-script/lan\-guage\_server/gdscript\-\_language\-\_server.cpp} includes \textit{editor\_log.h}) and VFX (\textit{mo\-dules/light\-map\-per\_\-cpu/light\-mapper\_cpu.cpp} includes \textit{editor\_settings.h}).

\begin{figure*}
\center
\includegraphics[width=1.4\columnwidth]{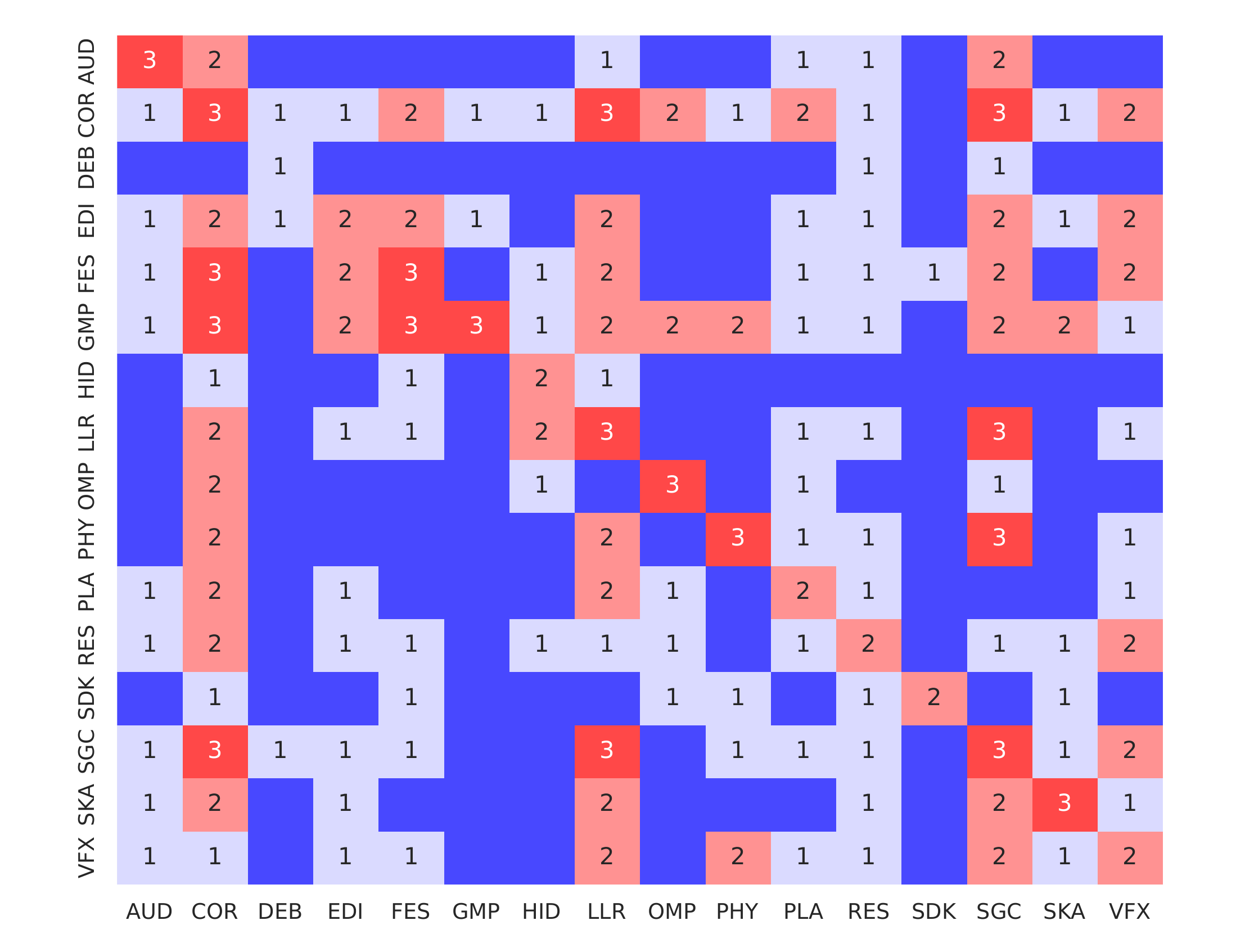}
\caption{Subsystem coupling heatmap for the selected game engines.}
\label{fig:heatmap-all}
\end{figure*}

In Urho3D, the COR subsystem has the highest in-degree, with four incoming and five outgoing edges. We were surprised, however, to find out that besides providing functionality, Urho3D's and Cocos2d-x COR also depend on other subsystems. They relate to graphics-related subsystems, such as FES, LLR, or VFX, for the following reasons:

\begin{itemize}
\item \textbf{Performing initialisation.} In Urho3D, \textit{Engine.cpp} includes \textit{Graphics.h} (LLR) to initialise the graphics and rendering subsystems. The graphics subsystem always starts up unless the engine is run in headless mode.

\item \textbf{Accessing debug information.} In Cocos2d-x, \textit{CCConsole\-.cpp} contains functions to write in a command line interface, which includes \textit{CCTextureCache.h} (LLR) and \textit{CCScene.h} (VFX) to print debugging information about the scene tree and texture cache to the console.
\end{itemize}

\subsubsection{Coupling Among Subsystems}
\label{sec:discussion-among-subs}

Besides analysing the game engines case by case, we want to understand what are the trends for game engine architecture in a broader sense. While we are aware that our selected engines may not represent the entire market and range of use cases of this kind of system, we believe that looking at the way subsystems relate throughout different systems may give us insights into how this kind of system should be built. Seeking to obtain such insights with regard to coupling, we counted the number of times each subsystem couples with each other on each selected engine. The result of this analysis is a matrix where both lines and columns represent subsystems. Zero represents no coupling, while one represents coupling exists for a given game engine. By summing up these matrices and colouring the highest values, we produced a heatmap (\autoref{fig:heatmap-all}).

In the x-axis of the heatmap, we can see how many times a subsystem includes another in all selected game engines. On the other hand, in the y-axis, we can see how many times a subsystem is included by another. For example, if we take the first line from the top, we can observe the AUD subsystem includes files from itself in three game engines, includes files from COR in two game engines, and so on.

The sum of values on each column (included by) and row (includes) is shown on \autoref{tab:freq-inc}. We put the top four subsystems in the ``included by'' column in the centre of each architectural map from \autoref{fig:arch-maps}. Since these subsystems are represented by nodes with many edges, putting them in the centre makes visualisation easier.

In the heatmap's central diagonal, we observe that not all values equal three. This happens because not all subsystems were detected in all game engines, and therefore not all self-include three times. This is the case for seven subsystems: DEB, EDI, HID, PLA, RES, SDK and VFX.

\begin{table}[ht]
\centering
\begin{tabular}{rlrlr}
\textbf{\#}             & \multicolumn{2}{l}{\textbf{Included By}} & \multicolumn{2}{l}{\textbf{Includes \phantom{000}}} \\ \hline
\multicolumn{1}{r|}{1}  & COR       & \multicolumn{1}{r|}{31}      & GMP                & 26               \\
\multicolumn{1}{r|}{2}  & SGC       & \multicolumn{1}{r|}{27}      & COR                & 25               \\
\multicolumn{1}{r|}{3}  & LLR       & \multicolumn{1}{r|}{26}      & FES                & 19               \\
\multicolumn{1}{r|}{4}  & VFX       & \multicolumn{1}{r|}{17}      & SGC                & 19               \\
\multicolumn{1}{r|}{5}  & FES       & \multicolumn{1}{r|}{16}      & EDI                & 18               \\
\multicolumn{1}{r|}{6}  & RES       & \multicolumn{1}{r|}{15}      & LLR                & 15               \\
\multicolumn{1}{r|}{7}  & PLA       & \multicolumn{1}{r|}{14}      & RES                & 15               \\
\multicolumn{1}{r|}{8}  & EDI       & \multicolumn{1}{r|}{13}      & VFX                & 15               \\
\multicolumn{1}{r|}{9}  & AUD       & \multicolumn{1}{r|}{12}      & PHY                & 13               \\
\multicolumn{1}{r|}{10} & SKA       & \multicolumn{1}{r|}{11}      & SKA                & 13               \\ \hline
\end{tabular}
\caption{Subsystems by include frequency.}
\label{tab:freq-inc}
\vspace{-0.30cm}
\end{table}
By inspecting the heatmap from left to right, we can see four columns are clearly highlighted: COR, LLR, SGC and VFX, indicating they are the subsystems most often included by others. As described in Gregory's architecture, COR files are ``useful software utilities'' \cite[p.~39]{gregory_game_2018} used throughout the system, and therefore it is no surprise that it is included by almost all subsystems. 

Video games are highly dependent on visuals, and graphics are a cross-cutting concern. Therefore it is also no surprise that so many subsystems depend on LLR and SGC. The reasons for this dependency only become clear when we look at the code. The most common example can be seen in Cocos2d-x: the FES subsystem, responsible for drawing UI elements, includes the renderer \textit{CCRenderer.h}. Urho3D PHY and AUD subsystems include files from LLR to print debugging information. In Cocos2d-x, PHY includes \textit{CCScene.h} because it holds a representation of the physics simulation, so it needs to add/remove physics objects.

\section{Threats to Validity}
\label{sec:threats}
Firstly, we are aware the selected game engines may not be representative of all open-source game engines and the entire video game industry. While we used GitHub stars and forks as a metric to enable us to select systems which are relevant to the development community, we observed that the nature of the open-source game engine ecosystem itself poses challenges that do not exist in other kinds of systems. 

For example, many open-source front-end web development frameworks exist and are being used in large industry-grade projects. Open-source game engines, on the other hand, are not used by large game development companies, being Unreal Engine one of the few exceptions. However, Unreal Engine has a large number of files and complex subsystems which would deserve a dedicated study. 

In the context of subsystem selection, Gregory's ``Runtime Engine Architecture'' is our single architectural reference for game engine subsystems. We know that other authors have proposed different reference architectures.

During subsystem detection, we noticed some files and folders did not fit into any of the selected subsystems. Hence, we are aware our subsystem list is not exhaustive.

In this study, subsystem detection was performed manually by the first author only. As described in \autoref{sec:include-graph-generation}, we also resolved some include paths manually. Therefore, we are aware both of these steps may have inconsistencies and biases.

We used Moose for creating models of the selected game engine files, folders, and \textit{include} relations. We used its built-in tools to query the entities in this model and generate architectural models. Other, similar tools exist and could provide different models.

\section{Conclusion}
\label{sec:conclusion}

We proposed an approach of subsystem detection, model generation, and visualisation of architectural models of game engines. We answered the following questions on three open-source game engines, Cocos2d-x, Godot, and Urho3D:

\paragraph{RQ1: Which subsystems are present in game engines?} We detected all 16 selected subsystems \cite{gregory_game_2018} in the selected game engines, even though we did not detect every subsystem in every game engine. When we did not detect a subsystem, it was usually because its functionalities were provided by other subsystems (e.g., \textit{Visual Effects} and \textit{Low-Level Renderer}, \textit{Profiling and Debugging} and \textit{Core Systems}).

\paragraph{RQ2: Which subsystems are more often coupled with one another?} We observed that different subsystems play a core role in their system: in Cocos2d-x, it is \textit{Platform Independence Layer}; in Godot, \textit{Resource Management} and \textit{World Editor}; in Urho3D, \textit{Core Systems}. When looking at the aggregated data of all game engines, \textit{Core Systems}, \textit{Low-Level Renderer}, \textit{Visual Effects}, and \textit{Scene Graph/Culling Optimizations} are the subsystems most often included by others. They are all related to either \textit{Core Systems} or graphics. \textit{Core Systems} often includes graphics-related subsystems. 

\paragraph{RQ3: Why are these subsystems coupled with each other?} In Cocos2d-x and Godot, \textit{Platform Independence Layer} provides functionalities used throughout the system, such as scripting, configuration and file system management. In Godot, \textit{Resource Management} provides game object classes (e.g., animations, materials). \textit{Core Systems} are included by many others and frequently include graphic-related subsystems for initialisation and debugging.

By applying our approach, game engine developers can understand whether subsystems are related and share responsibilities and whether this sharing helps fulfil the objectives of the game engine. By generating architectural models and coupling heatmaps of their systems, developers can understand which subsystems are available, how they depend on each other, and which subsystems are centralising responsibilities. 

In future work, we intend to apply this approach to a larger set of game engines and subsystems. We will ask several developers to perform subsystem detection so as to decrease classification biases. We will improve the \textit{include} graph generation, aiming for fully automatic resolution of \textit{include} paths to increase analysis speed and accuracy. We will explore other software quality metrics, such as cohesion and complexity, and how they affect architectural understanding and system maintainability. 

\bibliographystyle{plainnat}
\bibliography{main.bib}

\end{document}